# Unveiling Topological Charges and Their Manipulation in Electronic Circuits

Zarko Sakotic[1*], Predrag Stankovic[1], Vesna Bengin[1], Alex Krasnok[2], Andrea Alú[3,4], and Nikolina Jankovic[1]

[1]*BioSense Institute-Research Institute for Information Technologies in Biosystems, University of Novi Sad, Dr Zorana Djindjica 1a, 21101, Novi Sad, Serbia*

[2]*Department of Electrical and Computer Engineering, Florida International University, Miami, FL 33174, USA*

[3]*Advanced Science Research Center, City University of New York, New York, NY 10031, USA*

[4]*Physics Program, Graduate Center, City University of New York, New York, NY 10016, USA*

Leveraging topological properties in the response of electromagnetic systems can greatly enhance their potential. Although the investigation of singularity-based electromagnetics and non-Hermitian electronics has considerably increased in recent years in the context of various scattering anomalies, their topological properties have not been fully assessed. In this work, we theoretically and experimentally demonstrate that non-Hermitian perturbations around bound states in the continuum can lead to singularities of the scattering matrix, which are topologically nontrivial and comply with charge conservation. The associated scattering matrix poles, zeros and pole-zero pairs delineate extreme scattering events, including lasing, coherent perfect absorption, and absorber-lasers. The presented framework enables a recipe for generation, annihilation and addition of these singularities in electric circuits, with potential for extreme scattering engineering across a broad range of the electromagnetic spectrum for sensing, wireless power transfer, lasing and thermal emission devices.

## 1. Introduction

The discovery and utilization of the topological properties of physical systems has shown an increasingly significant impact on science and technology in recent years [1-3]. The emergence of topological phenomena in wave systems has opened new research directions, such as immunity to disorder [4], one-way transport [5], and topological lasing [6]. Beyond the analogies with condensed matter systems in realizing topological photonic insulators, topological concepts have recently enabled nontrivial scattering and radiative photonic phenomena [7,8], giving rise to new opportunities. Within this context, bound states in the continuum (BIC) arising in periodic structures have emerged as a remarkable phenomenon [9], with a range of applications stemming from their topological nature, such as polarization control [10-12], Q-factor enhancements [13], unidirectional BICs [14], vortex lasers [15] and vortex beam generation [16]. More recently, the topological nature of BICs was also proposed in non-periodic, planar reflective structures using epsilon-near-zero (ENZ) and epsilon-near-pole (ENP) materials [17], leveraged also in acoustic [18] and nonreciprocal electromagnetic systems [19-20]. BICs in planar reflective systems have been shown to be associated with pairs of perfect absorption singularities and reflection phase vortices, which comply with topological charge conservation. Separately from BIC-related phenomena, the topological nature of perfect absorption and reflection zero singularities was also recently studied in metasurfaces for polarization control [20,21] and optical multilayers for sensing [23, 24], highlighting the importance of topological concepts in scattering phenomena.

Perfect absorption is a special case of a more general phenomenon, coherent perfect absorption (CPA) in multiple-port systems [25,26]. CPA is the time-reversed version of lasing at threshold – both of which are related to singularities of the scattering-matrix eigenvalues, i.e., zeros (CPA) and poles (lasing at threshold). In this context, systems obeying parity-time-symmetry (PT-symmetry) have been shown to enable peculiar solutions of Maxwell's equations supporting exceptional points (EPs) [27], CPA-lasing degeneracies [28,29], unidirectional invisibility [30], unidirectional spectral singularity [31], virtual perfect absorption [32], and transient PT-symmetry [33]. All these $S$-matrix singularities form the basis for a plethora of applications, including enhanced sensing [34-37], unconventional lasing [38], topological wireless power transfer [39, 40], efficient energy storage [41], and non-reciprocal microwave transport [42]. Moreover, the general wave nature of these phenomena enables them in different wave systems, such as elastic [43], optomechanical [44] and complex electromagnetic scattering systems [45]. More recently, it was shown that it is possible to achieve a CPA-EP or zero-EP with distinct, broadened absorption lineshape [46,47]. Although singularity-based research has considerably proliferated in recent years, their topological features have not been fully assessed. For instance, elucidating the connections between BICs and other singular points may further boost their manipulation, and enable novel functionalities for high-performance devices.

Here, we present a new outlook on scattering matrix singularities and BICs, and experimentally demonstrate their topological nature in two-port networks. We start by discussing a broad range of lossless, non-periodic two-port structures that support BICs, including optical and electronic systems. We then show that a non-Hermitian perturbation destroys the BIC, creating pairs of topologically protected $S$-matrix singularities whose nature depends on the type of perturbation. In particular, we show that loss, gain and PT-symmetric perturbations lead to the creation of pairs of CPA, laser, and CPA-laser degeneracies, respectively. Each of these conditions is characterized by a winding number in the phase of the $S$-matrix eigenvalues, and all of them comply with charge conservation. We theoretically and experimentally demonstrate their topological features using electric circuits, validating the general nature of the presented ideas and proposing a recipe to generate topological singular points. These findings lie at the intersection of topology, non-Hermiticity, BICs and singular responses, and they may facilitate the design of new sensors, thermal emitters, wireless power transfer systems, lasers, and other electromagnetic devices.

## 2. BICs and topological *S*-matrix singularities

The conceptual sketch of topological charge creation from BICs is shown in Fig. 1(a): a BIC is transformed into a pair of topological charges corresponding to CPA (CPA-laser) states when introducing a defect in the system characterized by loss (PT-symmetry). Due to the ubiquitous wave nature of BICs, which can be realized across the electromagnetic spectrum in different platforms, we analyze three analogous systems where the proposed topological features can be observed. In this way, we aim at highlighting the general nature and wide applicability of the proposed framework. We start the discussion with the general two-port optical system shown in Fig. 1(b) – a dielectric cavity created between two identical resonant layers, representing a generalized Fabry-Perot-type photonic system. In this example, the permittivity of the resonant layers is assumed to have either an ENZ (plasma) or an ENP (Lorentzian) resonance. Such resonances are commonly found in isotropic, anisotropic and 2D materials, such as InSb, ITO, SiC, α-MoO$_3$, hBN. However, they can also be tailored at the frequency of interest using metamaterials and metasurfaces [44,45,48] – an example of an analogous metasurface system is sketched in Fig. 1(c). Neglecting loss for the moment, if one of the cavity resonances spectrally coincides with one of the material or metasurface resonances, a perfectly trapped mode or BIC can be induced [17, 52]. This feature can be attributed to waves experiencing a hard-wall boundary (perfect mirror) when the wave impedances of the top and bottom layers go to zero or infinity [17]. The resonant thickness of the dielectric, i.e., the BIC condition, depends on the permittivity $\varepsilon_d$ and incident angle $\theta$ as

$$d = \frac{n\lambda_r}{2} = \frac{nc}{2f_0\sqrt{\varepsilon_d - \sin^2\theta}}, \qquad n = 0,1,2\ldots \tag{1}$$

where $f_0$ represents the material or metasurface resonant frequency. As mentioned, this type of BIC can arise in a broad range of materials at different wavelengths, for example using SiC [50, 51], plasmonic metasurfaces [52], photonic crystals [53], and 2D-materials such as α-MoO$_3$ [54]. Two examples of optical systems supporting BICs, using α-MoO$_3$ layers and Ag-nanoparticle metasurfaces, are discussed in more detail in the Supplementary Materials. Although their topological nature in lossy one-port structures has recently been explored [17] and utilized for nonreciprocal thermal emission [19,20], these features have not yet been confirmed experimentally. Furthermore, their topological nature has not been discussed in two-port networks, where, apart from loss, PT-symmetry can also be studied.

An appealing platform to explore the topological features of BICs and related phenomena is formed by radio-frequency electronic circuits, due to feasibility, wide availability and cost-efficiency. To this end, in our study we propose the basic circuit, analogous to the proposed optical multilayer and metasurface structures, sketched in Fig. 1(d). For brevity, the following discussion will focus on this electric circuit, while the same analysis for optical systems is provided in the Supplementary Materials. The circuit has two ports and consists of two shunt $RLC$ resonant circuits separated by a transmission line of length $d$. This system is analogous to the optical structures under normal incidence when the resonant layers have an ENP resonance (Lorentzian) [17], i.e., the series impedance of the resonant layers has a minimum at resonance. In the circuit scenario, the polarization and incident angle degrees of freedom are not present. At the resonant frequency $\omega_0 = 1/\sqrt{LC}$, the impedance of the shunt lines is equal to $R$. When the transmission line length is

$$d = \frac{n\lambda_0}{2} = nc\pi\sqrt{LC}, \qquad n = 0,1,2 \dots, \qquad (2)$$

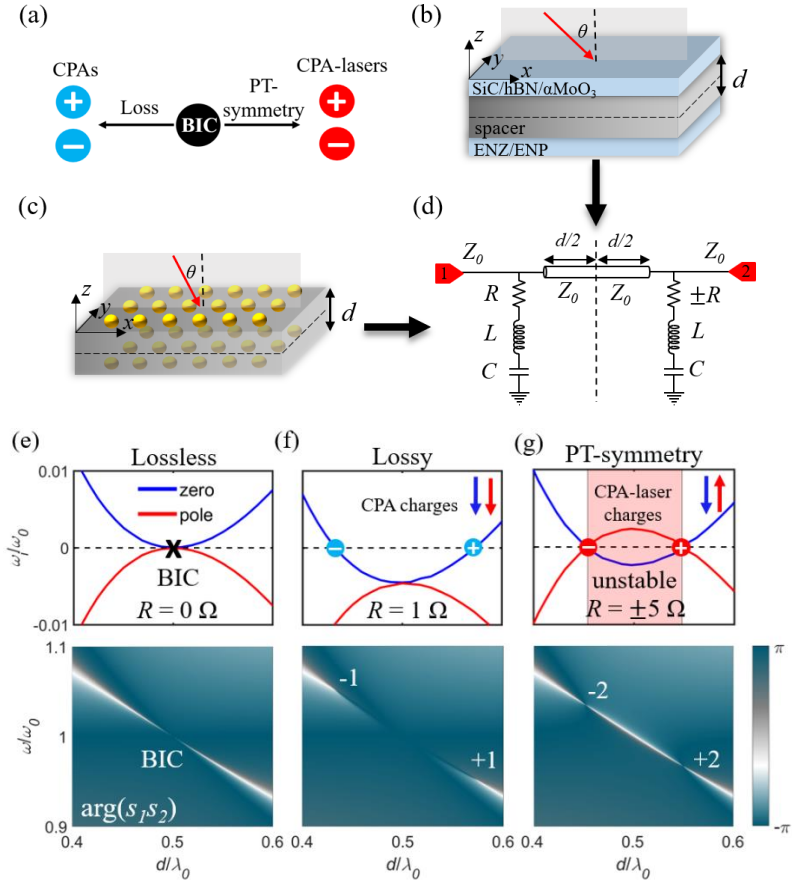

**Figure 1.** (a) Pairs of topological charges emerging from BICs in various physical systems. (b) Optical 3-layer system for a range of materials having ENZ or ENP resonances such as SiC, hBn, α-MoO$_3$. (c) Two plasmonic metasurfaces separated by a spacer. (d) Analogous electrical circuit supporting BICs. (e) $S$-matrix pole and zero dispersion for the circuit calculated in the $d$-$\omega_i$-$\omega_r$ space (upper panels) and phase of the multiplied eigenvalues in the $d$-$\omega_r$ plane (lower panels) for the lossless case $R=0$, (f) lossy case $R=1$ Ω, and (g) PT-symmetric case $R=\pm 5$ Ω.

and when *R* is ideally zero, the circuit supports a BIC – a Fabry-Perot mode perfectly trapped between two short circuits. In the optical case, this is equivalent to the Fabry-Perot dark mode of a dielectric cavity sandwiched between two perfect conductors. When loss, gain, or a PT-symmetric perturbation is added to the *RLC* circuits, the BIC is revealed to be a degeneracy of scattering matrix singularities that are topologically protected, as shown in Fig. 1(e-g) and analyzed further.

To analyze the BIC and the emerging charges, we use the complex frequency notation and the scattering matrix formalism. Specifically, we look for singularities (poles and zeros) of the scattering matrix eigenvalues in the complex frequency plane, which can describe the scattering process and emerging anomalies [50]. To visualize the charges appearing at real frequencies, we plot the phase of the multiplied eigenvalues of the *S*-matrix arg($s_1 s_2$), which contain all undefined phase points (vortices) associated with singular values, i.e., poles and zeros of the *S*-matrix eigenvalues where the phase is ill-defined. The polarity and winding number *q* of the charges can be found by integrating the phase accumulation counterclockwise in the parameter space around the charge $q = \frac{1}{2\pi} \oint d\phi$ [17]. The eigenvalues of the reciprocal two-port circuit are given by $s_{1/2} = t \pm \sqrt{r_l r_r}$, where transmission *t* and reflection coefficients $r_l$, $r_r$ are calculated using the ABCD matrix formalism [Supplementary Materials].

In the lossless case, the presence of BICs corresponds to a diverging phase resonance in the eigenvalue spectrum, as seen in the bottom panel of Fig. 1(e). In the complex frequency plane (upper panels), the dispersion of eigenvalue poles and zeros mirror each other due to time-reversal symmetry and Hermiticity, and they are parabolic around the resonant condition. Furthermore, the BIC is formed when these dispersion curves touch at the real-frequency axis [17]. When loss is added, Fig. 1(f), these dispersion curves move down along the imaginary frequency axis, inducing two real-frequency zeros of the scattering matrix, i.e., two CPAs. As loss is added, these points move in parameter space, but they cannot disappear unless they are annihilated with an oppositely charged singularity. The time-reversed scenario (not shown) entails adding gain in both resonators, hence shifting the curves up in the complex frequency plane and inducing two real-frequency poles, i.e., lasing conditions. It is worth noting that, by placing a perfect mirror in symmetry points (middle of the spacer) of the system, we obtain a one-port structure and CPA solutions become PA solutions, which were explored in detail in [17]. Lastly, we consider a PT-symmetric perturbation to the system, i.e., *RLC* tanks with positive and negative resistances, Fig. 1(g). Now, the zero-dispersion translates down, while poles translate up along the imaginary frequency axis. Remarkably, these dispersions intersect at the same real frequency, creating two CPA-laser points. The eigenvalue phase spectrum contains singularities with charges of $\pm 2$ in the phase, indicating the presence of both a pole and a zero. If PT-symmetry holds, CPA-laser points are topologically protected and move in parameter space for different |*R*| values.

Due to the translation of system poles in the upper complex half-plane around the BIC, the parametric region between two CPA-laser charges originating from the same BIC is inherently unstable. As we show later, these charges can disappear only through annihilation. However, their dispersion carries important implications for the system stability – an aspect often overlooked in theoretical studies of PT-symmetric systems. We note that a similar finding was previously reported in 1D periodic PT-symmetric systems [56], where PT-symmetric perturbations create so-called PT-BIC rings, with similar implications for stability [57], although the CPA-laser aspect was not discussed. The proposed theoretical framework reveals a fundamental connection between BICs and different scattering extrema, as well as the topological nature of the emerging singularities. Furthermore, the same phenomena were shown to arise in different electrical and optical systems, indictaing the ubiquity and wide applicability of the proposed theory, which we further discuss in the following and verify experimentally.

## 3. Experimental demonstration of CPA topological charge conservation in electric circuits

CPA points were shown to arise from BICs in pairs after adding loss to the *RLC* resonators in Fig. 1. This finding represents a generalization of the topological theory associated with perfect absorption points and BICs in single port structures proposed in [17]. To further generalize the discussion and gain more insight into these CPA states, we analyze the system shown in the inset of Fig. 2(a). We simplify the structure by analyzing two complex impedances $Z_R$ separated by a transmission line in a two-port network, where $Z_R$ represents the total complex impedance of the series *RLC* circuit at a fixed frequency $Z_R = R + j\omega L + 1/j\omega C$. Using the ABCD matrix formalism, we find the reflection/transmission coefficients and the analytical solutions for CPAs [Supplementary materials]. Namely, for a mirror-symmetric and reciprocal system, the zeros of the two eigenvalues are given by $s_{1/2} = t \pm r = 0$. The two solutions for $Z_R$ as a function of *d* give rise to symmetric and antisymmetric CPAs:

$$Z_{Rs} = \frac{Z_0}{2}\left(\cos(kd) + \sqrt{\cos^2(kd) + 2j\sin(kd)}\right), \quad (3)$$

$$Z_{Ra} = -\frac{Z_0}{2}\left(\cos(kd) - \sqrt{\cos^2(kd) + 2j\sin(kd)}\right), \quad (4)$$

where *k*=2*π*/*λ* is the wavenumber, and $Z_0$=50 Ω is the network characteristic impedance. For a fixed frequency *ω*=*ω₀*, Figs. 2 (a-b) show the CPA solutions for different transmission line lengths. When *d*=0 (no transmission line), the symmetric CPA exists for $Z_R$=50 Ω, while for the antisymmetric case, the solution requires that $Z_R$=0, i.e., the system supports an antisymmetric BIC and cannot absorb waves. As *d* increases, $Z_R$ solutions for CPA become complex, with opposite signs of the imaginary part for symmetric and antisymmetric cases. At $d$=$\lambda$/2, the system supports a symmetric BIC ($Z_R$=0) and an antisymmetric CPA ($Z_R$=50 Ω). These solutions periodically repeat as $d$ increases.

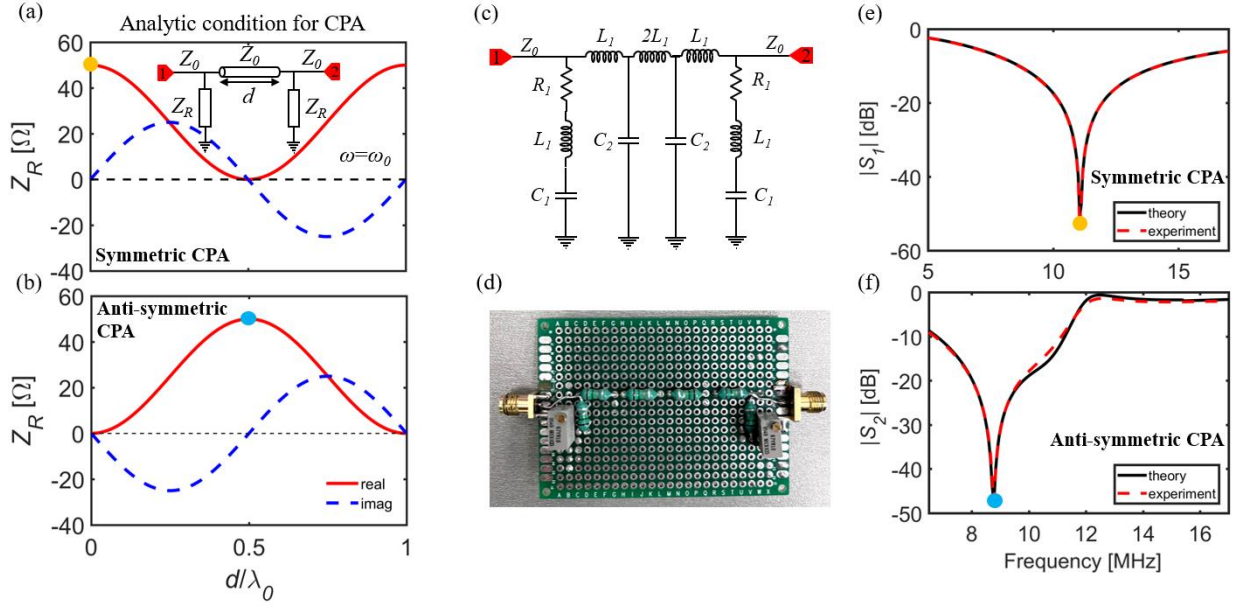

**Figure 2.** Analytical condition for (a) symmetric and (b) antisymmetric CPAs for the circuit model show in the inset of (a). (c) Schematic of the proposed RF-circuit. (d) Photograph of the realized circuit on a protoboard. (e) Comparison of the theoretical and measured eigenvalue amplitude for R=50 Ω, $L_1$=1000 nH, $C_1$=217 pF, and $d$=0 (no transmission line), showing a zero associated with symmetric CPA. (f) Antisymmetric CPA for $R$=50.5 Ω, $L_1$=$L_2$=1500 nH, $C_1$=$C_2$=217 pF which corresponds to $d$=$\lambda$/2. Network characteristic impedance is $Z_0$=50 Ω.

To confirm the existence of these CPAs and demonstrate their topological nature, we built an electric circuit consisting of two shunt *RLC* resonators separated by a double T-circuit, which plays the role of a transmission line and provides the required phase delay between two identical resonators [Supplementary Materials], Fig. 2(c). A photograph of the protoboard realization of the proposed electric circuit is shown in Fig. 2(d). Firstly, we validate the existence of the symmetric and antisymmetric CPAs and measure the response of the two circuits. To account for parasitic effects dominated by the series resistance of inductors, we add series resistors to each inductor in the circuit, which we evaluated to be ∼1 Ω at the measured frequencies. The first circuit has a resonant frequency of 10.8 MHz and no transmission line ($d$=0). With the total series resistance $R$=50 Ω (including parasitics) of the *RLC* tank, the circuit response is in perfect agreement with theory, Fig 2(e). The zero in the eigenvalue magnitude spectrum $|s_1|$=$|t+r|$ confirms the emergence of a symmetric CPA. The second circuit has a T-circuit between the two resonators with $R$=50.5 Ω, where the T-circuit response is identical to the one of a transmission line with length $d$=0.5$\lambda_0$ at the resonance frequency $f_0$ [Supplementary Materials]. The response displays an antisymmetric CPA with $|s_2|$=$|t-r|$ going to zero at the resonance frequency 8.82 MHz, Fig. 2(f), for which experimental and theoretical results are in excellent agreement. The resistance $R$ is slightly larger than the theoretically predicted $R$=$Z_0$=50 Ω due to the parasitic resistance in the *LC*-tanks of the transmission line.

In order to demonstrate the topological nature and charge conservation of these states, we experimentally show the annihilation of antisymmetric CPA charges using the same circuit in Fig. 2(d). To understand the dynamics of the annihilation process, we analyze the 2D parameter space consisting of frequency and capacitance $C_1$, exploring the zeros of $s_2$. In this parameter space, zeros (CPA charges) originating from neighboring BICs move towards each other for increasing values of $R$, Fig. 3(a). At the critical value of 50.5 Ω (middle panel of Fig. 3(a)), the charges meet and annihilate. This is evident when $R$ further increases, and the $s_2$ zeros associated with the CPAs disappear, as shown in the 4th and 5th panels of Fig. 3(a). The calculated trajectory of these states in the 3-D parameter space $f$-$R$-$C_1$ is shown in Fig. 3(b), where the third dimension is represented by color. By tuning the circuit elements, specifically $R$ and $C_1$, different CPA points can be accessed according to this dispersion.

To trace the dispersion of these charges in a real circuit, we modified the original circuit in Fig. 3(d) by adding additional, identical capacitances in the left and right $RLC$ tanks to yield the required capacitance. Furthermore, using identical potentiometers placed in both $RLC$ tanks, we can tune the resistance to access the desired CPA points. We measure the frequency response of the circuit at five different parameter points (values of $C_1$ and $R$ shown on panels in Fig. 3 (a)), represented by the colored horizontal lines and dots in Figs. 3(a) and 3(b), respectively. The theoretical and experimental results are in excellent agreement, Fig. 3(c). The zero-magnitude point associated with one antisymmetric CPA charge moves through the frequency space until it annihilates with an opposite charge for $R$=50.5 Ω. Thus, the topological nature of CPA states, their connection with BICs, and their annihilation have been explored and demonstrated for the first time. The presented recipe for their creation and annihilation can be extended to various wave systems, including the optical and infrared electromagnetic systems, as well as acoustic and mechanical systems, where topology plays an increasingly important role in the design of novel devices, including wireless power transfer, polarization control and thermal emission manipulation.

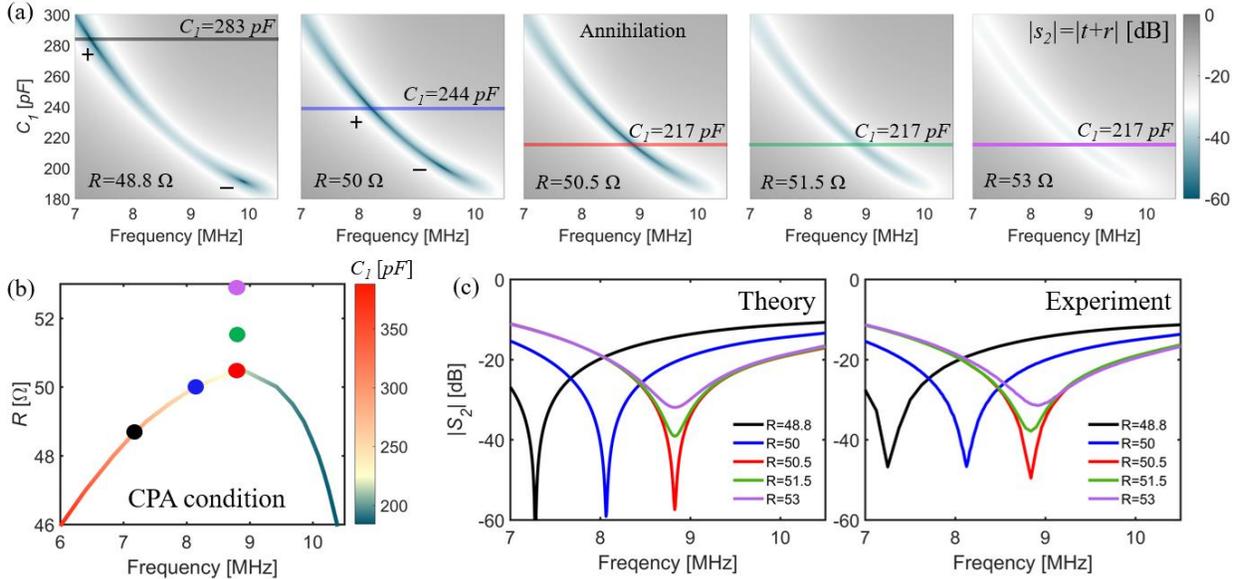

**Figure 3.** (a) Eigenvalue magnitude $|s_2|$ for five different values of $R$. Vertical colored lines represent the parametric points to be measured. (b) Numerically obtained CPA condition in the parameter space $f$-$R$-$C_1$, where $C_1$ is represented by the color. Colored dots represent different values for $R$-$C_1$ to be measured. (c) Theoretical and experimental obtained $|s_2|$ measured for 5 different parameter points, where the trajectory and eventual annihilation of the CPA is evident. Reactive components used are $C_2$=217 pF, $L_1$=$L_2$=1500 nH,

## 4. PT-symmetry and topological CPA-Laser charge

We next study a PT-symmetric RF circuit, showing a simple procedure to realize a topological CPA-laser degeneracy. We consider the system shown in Fig. 4(a), similar to the one analyzed in Fig. 2(a), but yielding positive and negative shunt resistance on different sides of the system – in such a way to obey PT-symmetry. A negative resistance can be realized using a negative impedance converter, which has been already used to explore PT-symmetry in circuit scenarios [58-60]. Alternatively, a recent demonstration has shown that building fully integrated PT-symmetric circuits is also possible [42], offering more flexibility in the design. Using the same ABCD matrix formalism as for the previous analysis, we compute the eigenmodes with real eigenfrequency, i.e., real-frequency poles of the scattering matrix eigenvalues [Supplementary materials]. The real and imaginary parts of the complex impedance $Z_R$ are

$$R = \frac{Z_0}{\sqrt{2}} \sqrt{\sin^2(kd) - \frac{\sin^2(2kd)}{8}} \tag{5}$$

$$X = -\frac{Z_0}{4} \sin(2kd) \tag{6}$$

Due to PT-symmetry, each real-frequency pole is degenerate with a zero, implying the emergence of CPA-laser states. Thus, Eqs. (5,6) constitute the conditions for CPA-laser states, Fig. 4(b). To show their usual features of absorbing and lasing at the same frequency, we plot the total output power for two CPA-laser points, emerging for $Z_R=20 \pm j12.4$ Ω at two different values of *d* for different phase inputs, Fig. 4(c). We plot the total outgoing or scattered power, defined as

$$P = |a_1 r_L + a_2 t|^2 + |a_2 r_R + a_1 t|^2 \tag{7}$$

where $a_1 = A_1 \cos(\omega t)$ and $a_2 = A_2 \cos(\omega t + \psi)$ represent input signals with amplitudes $A_{1/2}$ and with relative phase difference *ψ*. At the CPA-laser point, the output *P* is infinitely large, as there is a pole on the real frequency axis. However, exciting the system with the eigenvector corresponding to the zero of the *S*-matrix eliminates all outgoing power, and all the energy is absorbed, as the blue curves show in the graph. Additionally, we plot the phase of the eigenvalues in the inset, which clearly shows two points with charges of +2 and -2 associated with the CPA-laser condition. As the value of |*R*| increases, these charges move

closer to each other, until they meet at $d/\lambda_0=0.25$ and $R=50/\sqrt{2}$, and annihilate each other for $R>50/\sqrt{2}$, Fig. 4(d). As shown in the inset, the phase singularities disappear after annihilation.

A crucial aspect of active systems that has not yet been discussed is associated with their stability. As shown in Fig. 1(f) and briefly discussed above, PT-symmetric perturbations to the BIC creates an unstable parametric region between two emerging CPA-laser charges. The pole dispersion translates above the imaginary-frequency axis in the complex plane. Similarly, in Fig. 4(c) two CPA-laser states imply the emergence of points at which the pole dispersion crosses the real axis and moves to the upper complex frequency half-plane, indicating an unstable region (shaded red). As $|R|$ grows to $50/\sqrt{2}$ and beyond, the CPA-laser charges move to the same position and annihilate, at which point all poles are unstable, Fig. 4(d) and [Supplementary materials]. In the present analysis, thecircuit response is studied at fixed frequency, hence without considering the dispersion of the involved active circuitry. Additional discussions on this issue based on complex frequency analysis can be found in the Supplementary Materials. Interestingly, the case $|R|=50/\sqrt{2}$ was exploited in a similar circuit layout for enhanced sensing [61], indicating a possible route to experimentally explore the discussed features, operating close to instability Thus, the annihilation of CPA-laser charges would indeed be impossible to observe in a real system due to the inherent instability after annihilation. Nevertheless, the proposed concept presents an important connection between BICs, CPA-laser degeneracies, and topology. Our findings reveal a simple procedure to generate CPA-laser states, promising for next-generation sensors [37,61] and wireless power transfer systems [39-40, 62]. Additionally, we have formulated design guidelines to address stability, a crucial aspect for active and PT-symmetric systems.

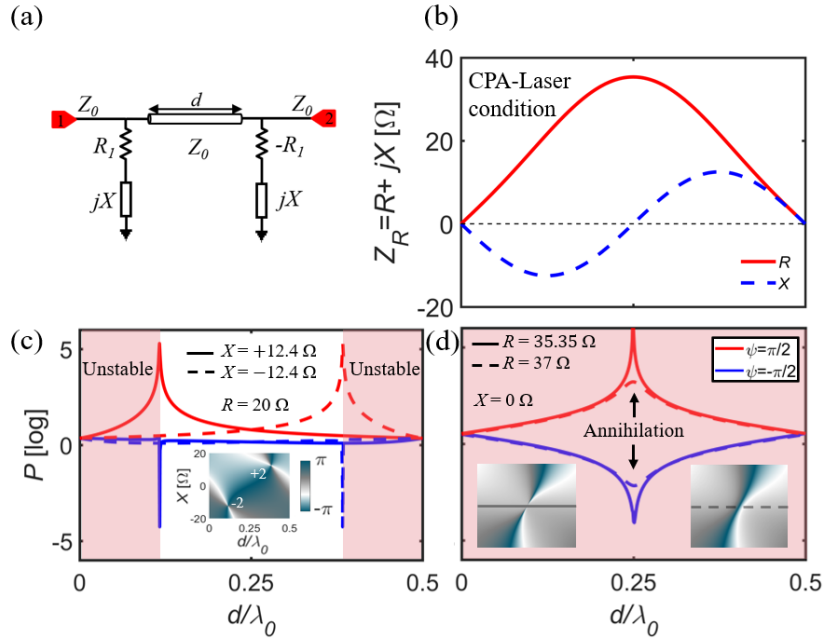

**Figure 4.** (a) PT-symmetric circuit. (b) Analytical condition for CPA-laser states at a fixed frequency. (c) Total outgoing power calculated at two CPA-Laser points with different excitation phases (red and blue lines). The inset shows the phase singularities associated with CPA-laser points. (d) P calculated at $R=50/\sqrt{2} \approx 35.355$ Ω where two CPA-Laser charges merge (phase $\arg|s_1 s_2|$ shown in the inset). For $R=37$ Ω, charges are destroyed as evident from the stripped line and right inset (no phase singularities).

## 5. Charge addition, CPA exceptional points and unidirectional spectral singularities

So far we have explored the conservation of topological charges through their annihilation. In this section, we explore the possibility of charge addition, which will entail merging of two CPA solutions, forming an exceptional point (EP) of zeros and enabling unusual physics. Traditionally, EPs emerge when two modes support degenerate eigenvalues and eigenvectors, with far-reaching consequences [27]. In the complex frequency space, an EP is represented by two merging poles [55]. In contrast, the time-reversed version of such condition is represented by merging two zeros, i.e., a CPA EP, which was recently theoretically and experimentally shown in optical microcavities [46,47]. Using the introduced topological features of CPAs, we demonstrate here that two CPA charges of the same polarity can merge to form an EP of zeros.

Using the passive version of the circuit in Fig. 5(a), two charges of same polarity can be brought together at the same real frequency. Due to the same polarity, CPA charges do not annihilate and have a total of +2 charge in phase space, while the analysis of the same system in the complex frequency plane reveals that there is an EP of both poles and zeros, Fig. 5(b). The existence of such a CPA EP in our system is not surprising: the T-circuit acts as a low pass filter and suppresses transmission for higher frequencies [Supplementary Materials]. In this scenario, when the transmission is eliminated the emerging CPAs will be identical to reflection zeros $s_{1/2} = t \pm \sqrt{r_l r_r} = 0$, $t = 0 \rightarrow r_l = r_r = 0$, and these zeros are co-located due to the mirror symmetry of the circuit.

Although this CPA EP appears trivial at first glance, a rather interesting feature arises when we analyze the PT-symmetric version of this circuit, which breaks mirror symmetry. By varying $R_2$ to 0, and then to -$R_1$ in the right side of the circuit, a unidirectional resonant transmission is achieved, which was first theoretically explored in [31]. Remarkably, the circuit is transparent from one side with zero reflection and full transmission, while from the other side, the circuit acts like a laser, Fig. 5(e). As mentioned earlier, negative resistance can be achieved using a negative-impedance converter configuration with an operational amplifier, as sketched in Fig. 5(a). To test our theoretical prediction obtained with an ideal negative resistor $R_2$, we simulated the response of the PT-symmetric circuit using a realistic model for the commercially available operational amplifier (Texas Instruments LMH6714) in the negative-impedance-converter configuration, which takes into account dispersion effects. This configuration provides an effective negative resistance, and the response is in excellent agreement with our theoretical prediction, Fig. 5(e). Although the T-circuit suppresses transmission, the peculiar physics of EPs at hand allows full transmission at the unidirectional spectral singularity, which can be exploited for wireless power transfer applications. This is expected, since one of the eigenvalues is zero, i.e., $s_{1/2} = t \pm \sqrt{r_l r_r} = 0$. Due to the singular values of the reflection coefficients ($r_l = 0$ and $r_r = \infty$) at the resonant frequency in this ideal PT-symmetric case, i.e., $|r_l| = 1/|r_r| = 0$, the amplitude of the transmission coefficient must be equal to 1. Notably, the system is stable, as the only pole near the real-frequency axis is associated with the discussed singularity and it can be explicitly controlled with $R_2$. These results reveal an essential connection between CPA EP and unidirectional spectral singularity. Furthermore, the unusual unidirectional character of the underlying EP might facilitate new solutions in wireless power transfer systems.

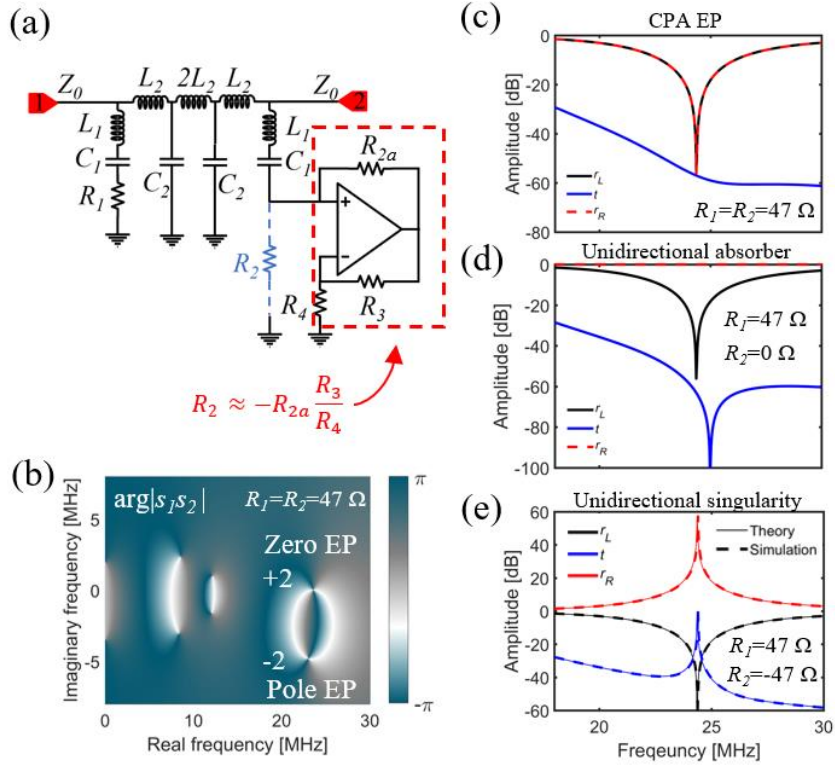

**Figure 5.** (a) Circuit with a negative impedance converter, $L_1 = L_2 = 1520$ nH, $C_1 = 27.8\ pF$, $C_2 = 217\ pF$. (b) Eigenvalue phase in the complex frequency plane where zero EP and pole EP are visible as $\pm 2$ charges. For a passive circuit $R_1=R_2 =47\ \Omega$, the zero-EP is at the real frequency axis forming a CPA-EP. (c) Reflection and transmission coefficients for lossy structure, same as (b). (d) Unidirectional absorber, (e) Unidirectional spectral singularity. For simulation purposes, we used $R_3=R_4=1000\ \Omega$, and $R_{2a}=49.5\ \Omega$ to obtain the required negative resistance at the desired frequency. We have also added an additional series inductance in the right circuit $L_p=100$ nH to cancel out the parasitic reactance added by the negative impedance converter and equalize the resonant frequencies of left and right resonant circuits.

## 6. Conclusion

Here we have explored and demonstrated the topological nature of *S*-matrix singularities, and their connection with BICs in two-port structures. Using a basic electric circuit, the trajectory and annihilation of CPA charges were experimentally demonstrated, with excellent agreement with theoretical predictions. Furthermore, the analysis of PT-symmetric circuits shows that topological CPA-laser charges emerge from BICs for PT-symmetric perturbations, revealing an important connection between topology, BICs and PT-symmetry. Using the proposed topological concepts, charge addition was also demonstrated, where two CPAs form an EP of zeros instead of annihilating each other. Due to the peculiarity of the CPA EP, a unidirectional absorber and unidirectional spectral singularity were also shown. The novel outlook on BICs and related singularities enable a simple procedure to leverage their topological properties, which can be applied to design singularity-based devices. Although demonstrated with electric circuits, these findings relate to materials and systems across the electromagnetic spectrum. We expect them to impact thermal emission, wireless power transfer, sensing and lasing engineering.

**METHODS**

For the experimental circuit realization, we have solderered standard *RLC* components onto a circuit protoboard. We used surface mount (SMD) 0805 capacitors and resistors, and fixed value axial inductors. For the variable resistor, we used BI-9549 10 Ohm Multiturn trimmer. Input/output ports were created by mounting two SMA connectors on the edges of the circuit board. Circuits were measured in the range 5-30 MHz with Vector Network Analyzer – AGILENT E5071C, and full scattering-matrix parameters were extracted from two-port measurements.

**ACKNOWLEDGEMENTS**

The work described in this paper is conducted within the project NOCTURNO, which receives funding from the European Union's Horizon 2020 research and innovation programme under Grant No. 777714, as well as the ANTARES project that has received funding from the European Union's Horizon 2020 research and innovation program GA 739570. The work was also partially supported by the Department of Defense, the Air Force Office of Scientific Research, the National Science Foundation and the Simons Foundation.

**AUTHOR CONTRIBUTION**

Z.S. conceived the idea for the work and performed the analytical and numerical analysis. Z.S. and P.S conducted the experimental part. Z.S. wrote the manuscript. A.K, N.J., V.B. and A.A. contributed to the manuscript and supervised the project.

# Supplementary material: Unveiling Topological Charges and Their Manipulation in Electronic Circuits

Zarko Sakotic[1*], Predrag Stankovic[1], Vesna Bengin[1], Alex Krasnok[2], Andrea Alú[3,4], and Nikolina Jankovic[1]

[1]BioSense Institute-Research Institute for Information Technologies in Biosystems, University of Novi Sad, Dr Zorana Djindjica 1a, 21101, Novi Sad, Serbia

[2]Department of Electrical and Computer Engineering, Florida International University, Miami, FL 33174, USA

[3]Advanced Science Research Center, City University of New York, New York, NY 10031, USA

[4]Physics Program, Graduate Center, City University of New York, New York, NY 10016, USA

## S1. Topological charge creation in optical structures

As mentioned in the main text, the effects of BIC splitting into topological CPA and CPA-laser charges can be realized in many different systems. To substantiate the claimed ubiquity of the proposed phenomenon, we analyze two examples of optical systems – a tri-layer structure consisting of two α-MoO$_3$ layers separated by a dielectric spacer, and a pair of Ag-nanosphere metasurfaces, also separated by a dielectric spacer. Firstly, we analyze the optical tri-layer – we consider normally incident wave polarized in the y-direction, Fig. S1 (a). The permittivity of the material along *y*-direction is shown in Fig. S1 (b), according to [1]. The material has a Lorentzian or epsilon-near-pole (ENP) resonance at the transverse optical phonon frequency $\omega_{TO}$, which is necessary for obtaining the BIC condition [2]. If losses in the material are artificially turned to zero, this structure supports a BIC, Fig. S1 (c, left). The spacer thickness BIC condition is given by equation (1) in the main text. Now, if the real losses are "turned back" on, the BIC is split into two zeros of the eigenvalue i.e., two CPAs, Fig. S1(c, middle and right). Further manipulation and annihilation of these charges is possible with different thicknesses of the α-MoO$_3$ layers, similar to the discussions in our previous work [2]. Finally, we analyze the PT-symmetric version of the system, where two CPA-laser charges emerge around the BIC, Fig. S1 (d). We note here that the same phenomena can be obtained with any materials with ENZ or ENP resonances, such as SiC, hBN, or InAs, to name a few.

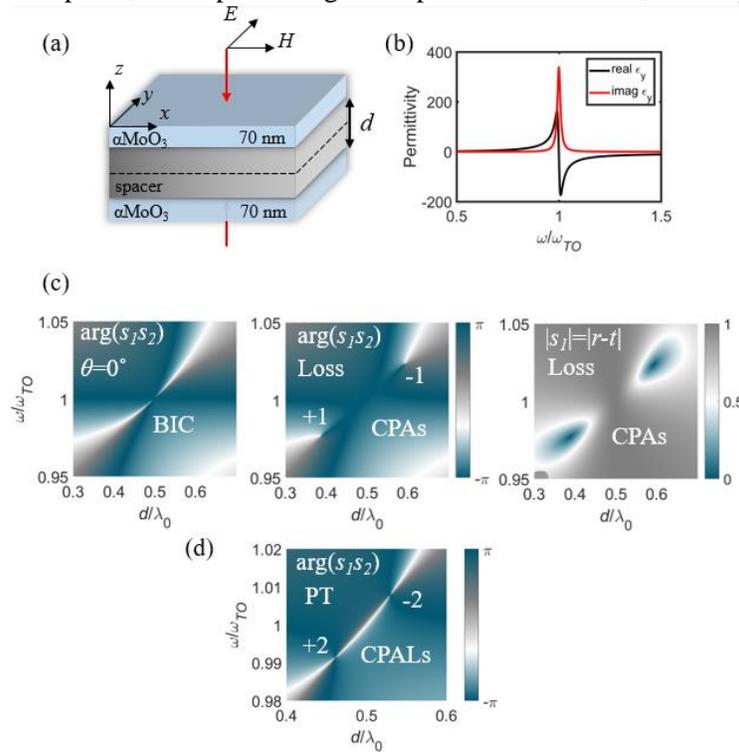

Figure S1. (a) Optical tri-layer structure with normally incident, y-polarized monochromatic wave, with dielectric permittivity $\varepsilon_d=5$. (b) Permittivity function $\varepsilon_y$ of α-MoO$_3$. (c) Eigenvalues phase for lossless (left) and real (middle) α-MoO$_3$, where BIC and CPAs are visible. Eigenvalue $|s_1|=|r-t|$ amplitude (right). (d) PT-symmetric structure with CPA-laser charges emerging from BIC.

On the other hand, the Lorentzian resonance can be effectively induced by metallic or dielectric metasurfaces utilizing the electric dipole resonance. To that end, we analyze a pair of silver nanoparticle metasurfaces as shown in Fig. S2(a), and previously analyzed in [3]. The diameter of the nanospheres considered is *r*=15 nm, while the pitch is *l*=35 nm. In this regime, only specular reflection is allowed, while no diffraction channels are open [3]. The effective conductivity induced by the metasurfaces is shown in Fig. S2 (b), which was calculated according to the analytical solution given in [3]. As shown in Fig. S2 (c,d), the previously discussed phenomenon of BIC splitting into two CPA or CPA-laser solutions is shown. We thus note that the discussed topological nature of BICs and *S*-matrix singularities concerns a wide range of materials and metasurface systems. Furthermore, the same effects are expected to be found in acoustics and mechanics, given the ubiquitous wave nature of the proposed framework.

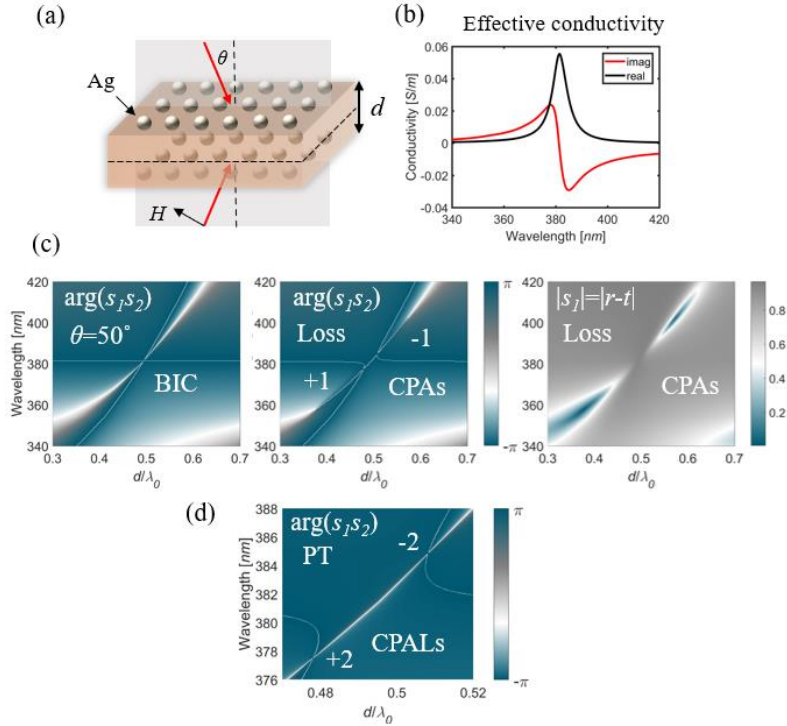

Figure S2. (a) Two Ag-nanoparticle metasurfaces separated by a spacer, with obliquely incident, y-polarized monochromatic wave. (b) Effective metasurface conductivity. (c) Eigenvalues phase for lossless (left) and real (middle) Ag metasurface where BIC and CPAs are visible. Eigenvalue $|s_1|=|r-t|$ amplitude (right). (d) PT-symmetric structure with CPA-laser charges emerging from BIC.

## S2. CPA condition

Here we derive the analytical solution for CPAs shown in the main text in equations (3) and (4). Using the ABCD matrix formalism [4], the total matrix of the circuit analyzed in Fig. S3 (inset of Fig. 2(a) in the main text) can be written as:

$$M_T = M_R M_{TL} M_R = \begin{bmatrix} 1 & 0 \\ \frac{1}{Z_R} & 1 \end{bmatrix} \begin{bmatrix} \cos(kd) & jZ_0 \sin(kd) \\ \frac{j}{Z_0} \sin(kd) & \cos(kd) \end{bmatrix} \begin{bmatrix} 1 & 0 \\ \frac{1}{Z_R} & 1 \end{bmatrix} =$$

$$= \begin{bmatrix} \cos(kd) + j\frac{Z_0}{Z_R} \sin(kd) & jZ_0 \sin(kd) \\ \frac{2}{Z_R} \cos(kd) + \frac{j}{Z_0} \sin(kd) \left(\frac{Z_0^2}{Z_R^2} + 1\right) & \cos(kd) + j\frac{Z_0}{Z_R} \sin(kd) \end{bmatrix} = \begin{bmatrix} A_T & B_T \\ C_T & D_T \end{bmatrix}$$

(s1)

The reflection and transmission coefficients are then calculated as:

$$r = \frac{A_T + B_T/Z_0 - C_T Z_0 - D_T}{A_T + B_T/Z_0 + C_T Z_0 + D_T} \tag{s2}$$

$$t = \frac{2}{A_T + B_T/Z_0 + C_T Z_0 + D_T}. \tag{s3}$$

We next set the eigenvalues to zero $s_{1/2} = t \pm r = 0$. For the anti-symmetric case we have $s_1 = t - r = 0$, and we solve for $Z_R$:

$$2Z_R^2 + 2Z_0 \cos(kd) Z_R + jZ_0^2 \sin(kd) = 0, \tag{s4}$$

which gives two complex solutions for $Z_R$:

$$Z_{Ra\ 1/2} = Z_0 \frac{-\cos(kd) \pm \sqrt{\cos^2(kd) - 2j\sin(kd)}}{2}. \tag{s5}$$

Only one of these solutions gives a positive real part for the impedance $Z_R$, so we choose that solution (plus sign). On the other hand, symmetric case $s_1 = t + r = 0$ leads to two solutions:

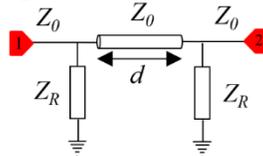

Figure S3. Transmission line circuit separated by two identical complex impedances that supports BIC and CPAs.

$$Z_{Rs\,1/2} = Z_0 \frac{\cos(kd) \pm \sqrt{\cos^2(kd) + 2j\sin(kd)}}{2} \tag{s6}$$

Similarly, we choose the plus sign to obtain the positive real part solution for $Z_R$. Equations s5 and s6 lead to final equations (3-4) and Figure 2(a-b) in the main text.

## S3. CPA-laser condition

To find CPA-laser solutions of the PT-symmetric circuit from Fig. 4 in the main text, we need to derive the conditions for zeros and poles of the eigenvalues $s_{1/2} = t \pm \sqrt{r_l r_r}$. Since the system is PT-symmetric, any real-frequency pole will be collocated with a real-frequency zero, thus only finding the pole dispersion is sufficient. The left (loss) and right (gain) resonators can be represented as complex admittances $Y_1=1/Z_1$ and $Y_2=1/Z_2$, such that PT-symmetry holds $Z_1 = -Z_2^*$, where the complex impedances are given by $Z_1 = R + jX$, $Z_2 = -R + jX$, and $R$ and $X$ are real numbers $R, X \in \mathbb{R}$. Since the reflection coefficients are different from different ports, we calculate the left and right total ABCD matrices, as well as reflection and transmission coefficients, as:

$$M_{Tl} = M_{loss} M_{TL} M_{gain} = \begin{bmatrix} 1 & 0 \\ Y_1 & 1 \end{bmatrix} \begin{bmatrix} \cos(kd) & jZ_0 \sin(kd) \\ \frac{j}{Z_0}\sin(kd) & \cos(kd) \end{bmatrix} \begin{bmatrix} 1 & 0 \\ Y_2 & 1 \end{bmatrix} =$$

$$= \begin{bmatrix} \cos(kd) + jZ_0 Y_2 \sin(kd) & jZ_0 \sin(kd) \\ \cos(kd)(Y_1 + Y_2) + \frac{j\sin(kd)}{Z_0}(1 + Z_0^2 Y_1 Y_2) & \cos(kd) + jZ_0 Y_1 \sin(kd) \end{bmatrix} \tag{s7}$$

$$= \begin{bmatrix} A_{Tl} & B_{Tl} \\ C_{Tl} & D_{Tl} \end{bmatrix}$$

$$M_{Tr} = M_{gain} M_{TL} M_{loss} = \begin{bmatrix} 1 & 0 \\ Y_2 & 1 \end{bmatrix} \begin{bmatrix} \cos(kd) & jZ_0 \sin(kd) \\ \frac{j}{Z_0}\sin(kd) & \cos(kd) \end{bmatrix} \begin{bmatrix} 1 & 0 \\ Y_1 & 1 \end{bmatrix} =$$

$$= \begin{bmatrix} \cos(kd) + jZ_0 Y_1 \sin(kd) & jZ_0 \sin(kd) \\ \cos(kd)(Y_1 + Y_2) + \frac{j\sin(kd)}{Z_0}(1 + Z_0^2 Y_1 Y_2) & \cos(kd) + jZ_0 Y_2 \sin(kd) \end{bmatrix} \tag{s8}$$

$$= \begin{bmatrix} A_{Tr} & B_{Tr} \\ C_{Tr} & D_{Tr} \end{bmatrix}$$

$$r_l = \frac{A_{Tl} + B_{Tl}/Z_0 - C_{Tl}Z_0 - D_{Tl}}{A_{Tl} + B_{Tl}/Z_0 + C_{Tl}Z_0 + D_{Tl}} = \frac{q_l}{p_l} = \frac{q_l}{p} \tag{s9}$$

$$r_r = \frac{A_{Tr} + B_{Tr}/Z_0 - C_{Tr}Z_0 - D_{Tr}}{A_{Tr} + B_{Tr}/Z_0 + C_{Tr}Z_0 + D_{Tr}} = \frac{q_r}{p_r} = \frac{q_r}{p} \tag{s10}$$

$$t_r = \frac{2}{A_{Tl} + B_{Tl}/Z_0 + C_{Tl}Z_0 + D_{Tl}} = \frac{2}{p} \tag{s11}$$

As required by reciprocity, the transmission coefficient is equal from both sides, i.e., the denominators are equal in equations (s9-s11) – $p_l=p_r=p$. For brevity purposes, we write the numerators of reflection coefficients as $q_l$ and $q_r$, and denominator as $p$. The eigenvalues are then given by:

$$s_{1/2} = t \pm \sqrt{r_l r_r} = \frac{2 \pm \sqrt{q_l q_r}}{p} \tag{s12}$$

The pole condition requires that the denominator $p$ is equal to 0, which gives the following equation:

$$(2 + Z_0(Y_1 + Y_2))e^{jkd} + j\sin(kd) Z_0^2 Y_1 Y_2 = 0. \tag{s13}$$

When $Y_1 = 1/Z_1 = 1/(R + jX)$ and $Y_2 = 1/(-R + jX)$ are inserted in equation (s13), the following equation can be obtained:

$$R^2 + X^2 + jZ_0 X = \frac{Z_0^2}{2} \frac{1}{1 - j\cot(kd)}. \tag{s14}$$

As $R$ and $X$ are real numbers, we can equate the real and imaginary parts of the left and right sides of the equations(s14) as:

$$Z_0 X = imag\left(\frac{Z_0^2}{2} \frac{1}{1 - j\cot(kd)}\right), \tag{s15}$$

$$R^2 + X^2 = real\left(\frac{Z_0^2}{2} \frac{1}{1 - j\cot(kd)}\right). \tag{s16}$$

Since $Z_0$ is a real number, we can write the solution for the imaginary part of the complex impedance as:

$$X = \frac{Z_0}{2} imag\left(\frac{1}{1 - j\cot(kd)}\right). \tag{s17}$$

After some trigonometric manipulation, this is further simplified to:

$$X = -\frac{Z_0}{4}\sin(2kd). \tag{s18}$$

Similarly, equation (s16) is simplified to:

$$R = \frac{Z_0}{\sqrt{2}}\sqrt{\sin^2(kd) - \frac{\sin^2(2kd)}{8}}. \tag{s19}$$

The last two equations represent the complete CPAL solution shown in the main text.

## S4. Stability and complex frequency analysis

As mentioned in the main text, stability is an important aspect of the PT-symmetric systems analyzed in this paper. As discussed with Fig. 1 in the main text, PT-symmetric perturbation separates the BIC into a pair of CPA-laser states, which act like topological charges of $\pm 2$. Due to the pole dispersion crossing to the upper complex half-plane, the region between two CPA-laser charges originating from the same BIC is necessarily unstable. We continued the discussion in Fig. 4, where annihilation of CPA-laser states was shown. Here we additionally discuss the same circuit with the imaginary frequency dimension, and we plot the dispersion of $S$-matrix poles and zeros (red and blue lines), Fig. S4.

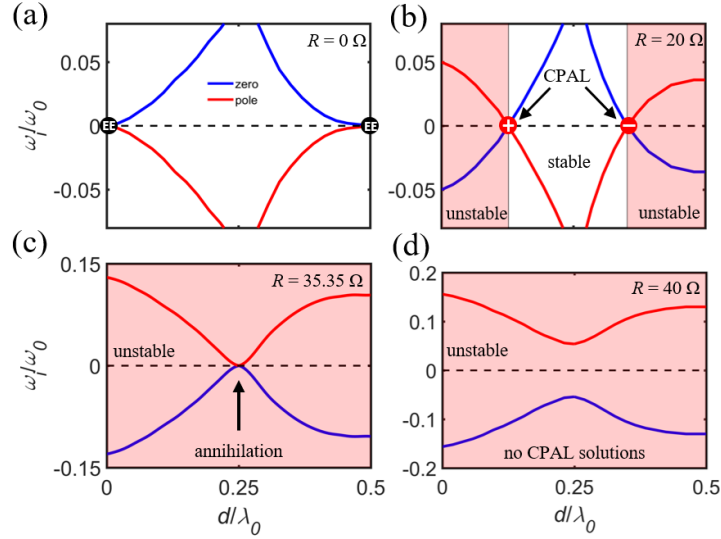

Figure S4. S-matrix pole and zero dispersion of the circuit from Fig. 4 in the main text. Solutions were found in the parameter space of $d$-$\omega_i$-$X$ for four different $|R|$ values (a-d). Red-shaded parts of the graphs represent the unstable regions.

In the lossless case ($R=0$), the BICs are located at 0 and 0.5 $d/\lambda_0$ transmission line thicknesses, Fig. S4 (a). As CPA-laser charges emerge out of neighboring BICs for $R>0$, the poles cross to the positive imaginary frequency space, indicating instability. As $R$ increases to $R=50/\sqrt{2}$ and the two charges are collocated (at annihilation point), the entire pole dispersion is in the upper complex half-plane, Fig. S4 (c), meaning that there is no stable solution in the entire parameter space. As $R$ increases further, there are no CPA-laser solutions left, poles move higher up, and the system remains unstable, Fig. S4 (d).

It should be noted that the fixed-frequency framework in which this analysis is done does not consider the dispersion of the negative resistance, which is unavoidable in real circuits due to causality. Although the creation, dispersion, and associated instabilities of CPA-laser degeneracies are thoroughly explained with this analysis, it should be used synergically with any dispersions present in real circuits to fully assess the system stability.

## S5. Analogy between transmission lines used in theoretical analysis and T-circuits used in realistic and experimental circuits

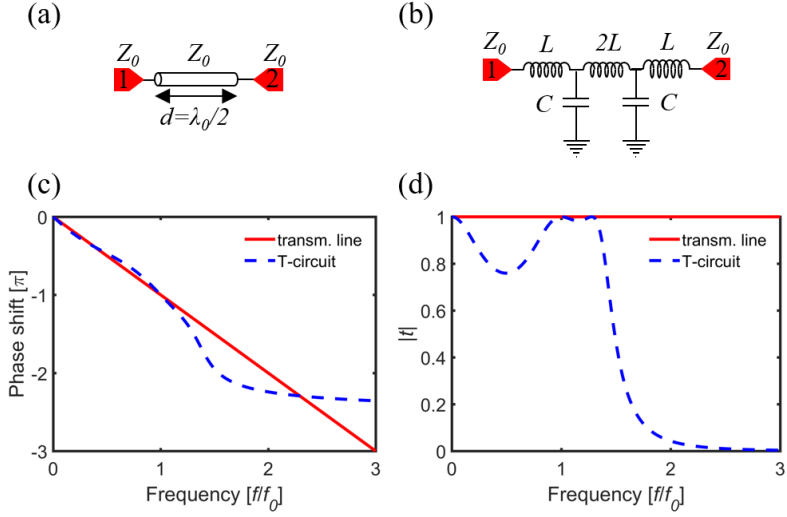

Figure S5. Phase and amplitude of the transmission coefficient comparison between the (a) transmission line and (b) double T-circuit.

In the theoretical analysis of CPA and CPAL states in the main text, we have used a two-port circuit consisting of two shunt *RLC* resonators separated by a transmission line, Fig. (1,2,4). For practical reasons and experimental purposes, we demonstrated the same effects using a double T-circuit consisting of multiple LC tanks instead of a transmission line, in Figs. (2,5) in the main text. Due to the large physical length of transmission lines at discussed frequencies ($\lambda_0/2 \approx 17$ m at $f_0$=8.8 MHz), a lumped element LC circuit can replace the transmission line as we show next. To validate this analogy, we plot the phase and amplitude response of the two circuits shown in Fig. S3. The double T-circuit in Fig. S5 (b) acts as a low-pass filter, whose response in the passband is very similar to that of the transmission line with length $d=\lambda_0/2$, and is most similar around $f=f_0$, where the topological properties were tested in real circuits.